\documentclass[conference]{IEEEtran}
  
\usepackage{amsmath,amssymb,amsfonts}

\usepackage{graphicx}
\usepackage{float}
\usepackage{subcaption}
\usepackage{tikz}
\graphicspath{{./images/}}

\usepackage{algorithm}
\usepackage{algpseudocode}
\usepackage{algorithmicx}
\usepackage{color}
\algrenewcommand\Require{\item[\textbf{Input:}]}

\usepackage{listings}
\def\BibTeX{{\rm B\kern-.05em{\sc i\kern-.025em b}\kern-.08em
    T\kern-.1667em\lower.7ex\hbox{E}\kern-.125emX}}

\usepackage{booktabs}
\usepackage{tabularx}
\usepackage{multirow}

\usepackage{comment}
\usepackage{enumitem}
\usepackage{soul}         % for \hl{}
\usepackage{ulem}         % for \uline{}
\usepackage{tcolorbox}
\usepackage{url}
\usepackage{balance}
\usepackage[hidelinks]{hyperref}

\newcommand{\tool}{{\sc SWIRL}}
\let\oldsout\sout
\renewcommand{\sout}[1]{\oldsout{#1}}

\begin{document}

\title{SWIRL: Interactive Sensemaking of Tool-Generated Warnings through Customized Summaries}

%\author{\IEEEauthorblockN{Anonymous Authors}
%\IEEEauthorblockA{Affiliation withheld for double-blind review}}

\author{\IEEEauthorblockN{Burak Yetiştiren}
\IEEEauthorblockA{\textit{Computer Science} \\
\textit{UCLA}\\
Los Angeles, CA, USA \\
burak@cs.ucla.edu}
\and
\IEEEauthorblockN{Hong Jin Kang}
\IEEEauthorblockA{\textit{The University of Sydney}\\
Sydney, Australia \\
hongjin.kang@sydney.edu.au}
\and
\IEEEauthorblockN{Miryung Kim}
\IEEEauthorblockA{\textit{Computer Science} \\
\textit{UCLA}\\
Los Angeles, CA, USA \\
miryung@cs.ucla.edu}
}

\maketitle

\begin{abstract}
Programmers using bug-finding tools often review their reported warnings one by one. Based on the insight that identifying recurring themes and relationships can enhance the cognitive process of searching for representations of a given problem space (i.e., \textit{sensemaking}), we propose \tool{}, which supports interpreting tool-generated warnings through interactive, customized summarization. With active feedback, \tool{} derives summary rules for grouping of related warnings on the fly. As users mark warnings as interesting or uninteresting, \tool{}'s rule inference algorithm surfaces common characteristics, highlighting structural similarities in containment, subtyping, invoked methods, accessed fields, and expressions.

We demonstrate \tool{} on real-world warnings generated from Infer and SpotBugs on two mature Java projects. In a within-subject user study, our participants articulated root causes for similar uninteresting warnings with more confidence when using \tool{}, compared to the baseline that lists individual warnings without customized summary rules. Among participants, we observed significant individual variation in desired grouping, reinforcing the need for individualized sensemaking. The simulation we conducted shows that {\tool}’s rule-level feedback expedites sensemaking, requiring only 11.8 interactions on average to align all inferred rules with a simulated user’s labels when combined with instance-level feedback, compared to 17.8 interactions when using instance-level feedback alone. Our evaluation suggests that \tool{}’s active learning-based summarization can enhance the sensemaking process of tool-generated warnings.
\end{abstract}



% ====================
% Main Sections
% ====================

\section{Introduction}
\begin{figure*}
\centering
    \includegraphics[width=\textwidth]{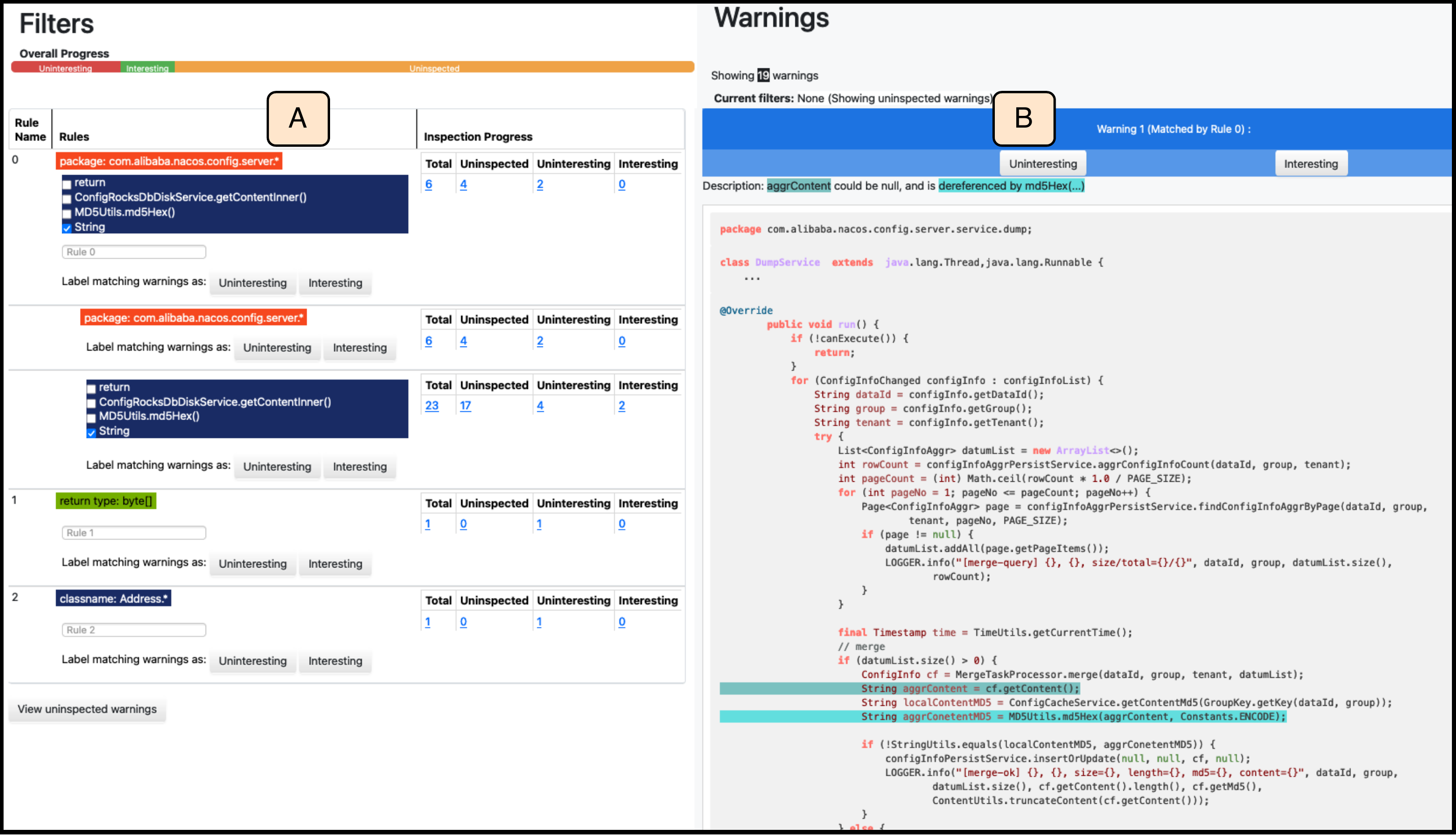}
    \caption{\tool{} guides users in \textit{sensemaking} of tool-generated warnings with inductive summary rules. Users can provide explicit feedback via (A) checkmarking code expressions or (B) highlighting code snippets. \tool{} then refines the summary rules interactively.} 
  \label{fig:swirl_buttons}
\end{figure*}

% large volume of results 
Bug-finding tools, such as Findbugs~\cite{ayewah2008using} and Infer~\cite{calcagno2015moving}, have been successful in detecting issues.
However, they also produce numerous uninteresting warnings that the developers do not want to investigate further~\cite{johnson2013don,christakis2016developers, Shen_Fang_Zhao_2011, designingUIs2021}. Users often struggle to make sense of these, or easily distinguish them from the interesting warnings and this difficulty has caused reluctance to adopt these bug-finding tools into a developer workflow~\cite{johnson2013don,christakis2016developers,zampetti2017open}. 
In our own analysis of the Alibaba Nacos open-source project~\cite{alibaba/nacos_2024}, we find that 43\% of the 58 \texttt{Null Pointer Dereference} warnings generated by Infer are never fixed, indicating that they are probably ignored by developers\footnote{based on our analysis in \url{https://github.com/UCLA-SEAL/SWIRL/tree/master/code/neverfixed_analysis}}. Evidently, the burden of sifting through a large number of tool-generated warnings has impeded the effective adoption of bug finding tools~\cite{johnson2013don, christakis2016developers,Shen_Fang_Zhao_2011, Smith_Do_Murphy-Hill_2020}.
% could erode in developer trust in tools by overwhelming them with warnings they are unable to sensemake.}
% no sense making
Existing research has focused on filtering or reprioritizing warnings~\cite{guo2023mitigating} but overlooked the problem of \textit{sensemaking}, defined as the essential task of searching for a representation or representations of the problem space or the underlying structure of a given set of information. In our use case scenario, we define sensemaking as \textit{making sense of related uninteresting (or interesting) warnings and identifying their common characteristics out of a pool of all warnings generated by a given bug-finding tool}. 

Unlike approaches that ask “Is this single warning a false positive or a true bug?”, sensemaking helps developers construct mental models for understanding the code and warnings reported on them.
% sense making
Inspired by the variational theory of learning~\cite{marton2014necessary}, where a user should form accurate conceptualizations by discerning variations of a phenomenon, we introduce SWIRL (\textbf{S}ensemaking \textbf{W}arnings with \textbf{I}nteractive \textbf{R}eview and \textbf{L}earning). \tool{} surfaces the commonalities and differences between warnings upon interactive feedback from a user. \tool{} empowers developers to create customized {flexible groups of warnings} by showing a customized scaffolding constructed with user feedback to examine related warnings together.

\tool{}'s interface is shown in Figure \ref{fig:swirl_buttons}.
Instead of viewing individual warnings one by one, using \tool{}, a user can formulate customized summaries of warnings by providing interactive feedback. This offers configurability, such that the summary rules are learned from user's labeling and code statement-level feedback on individual warnings. These summary rules reveal common characteristics based on code locations, type hierarchies, and similar code statements. Therefore, the user can easily make sense of similarities and dissimilarities between different inferred groups. 

\tool{} provides the following capabilities to improve sensemaking: 

\indent{\textbf{\textit{Feedback-driven Alignment.}}} \tool{} infers/re-infers summary rules, as a user  iteratively provides more feedback. This introduces a feedback loop. 
In this process, the warnings are grouped by rules, which is by no means filtering out any warnings, but assigning them to groups when applicable (when a warning matches an inferred group).

\tool{} employs inductive logic programming (ILP) to derive rules over containment, typing relations, and similar code expressions. These relationships capture the implicated code's scope, type hierarchy, and the API signature of a method where a warning is reported (e.g., the return type of the method, or the class fields that the method uses). This is useful since uninteresting warnings are often correlated by locality and organization of the implicated code~\cite{kremenek2004correlation}.
% \sout{For example, code located in the same package exposes HTTP endpoints related to unintended remote access~\cite{GitLab_2020}.
% As another example, a user may wish to group null pointer dereference warnings related to \texttt{getProperty("os.name")}, as it would return a non-null value at runtime~\cite{sysprop}. A user should be able to group warnings based on code cloning relationships~\cite{jiang2007deckard,kim2005empirical,saini2018oreo} or invocation of specific API  methods~\cite{van2020tailoring}.}

\indent{\textbf{\textit{Active Learning.}}} Active learning traditionally refers to learning settings in which the learner exercises some control over the data from which it learns, typically by selecting unlabeled instances and querying an oracle, such as a human annotator, for their labels~\cite{cohn1994active, settles2009active}. In our work, we use the term active learning in a broader, user-directed sense. Rather than having the learner select the instances to be labeled, we allow users to choose warning instances and provide labels based on their own sensemaking needs (interactive rule induction). The tool then incorporates these labels to iteratively induce and refine personalized grouping rules. Thus, our approach retains the iterative label--learn--update cycle of active learning, but places control over instance selection with the user rather than the learning algorithm.

\begin{figure}[!t]

    \includegraphics[width=0.5\textwidth]{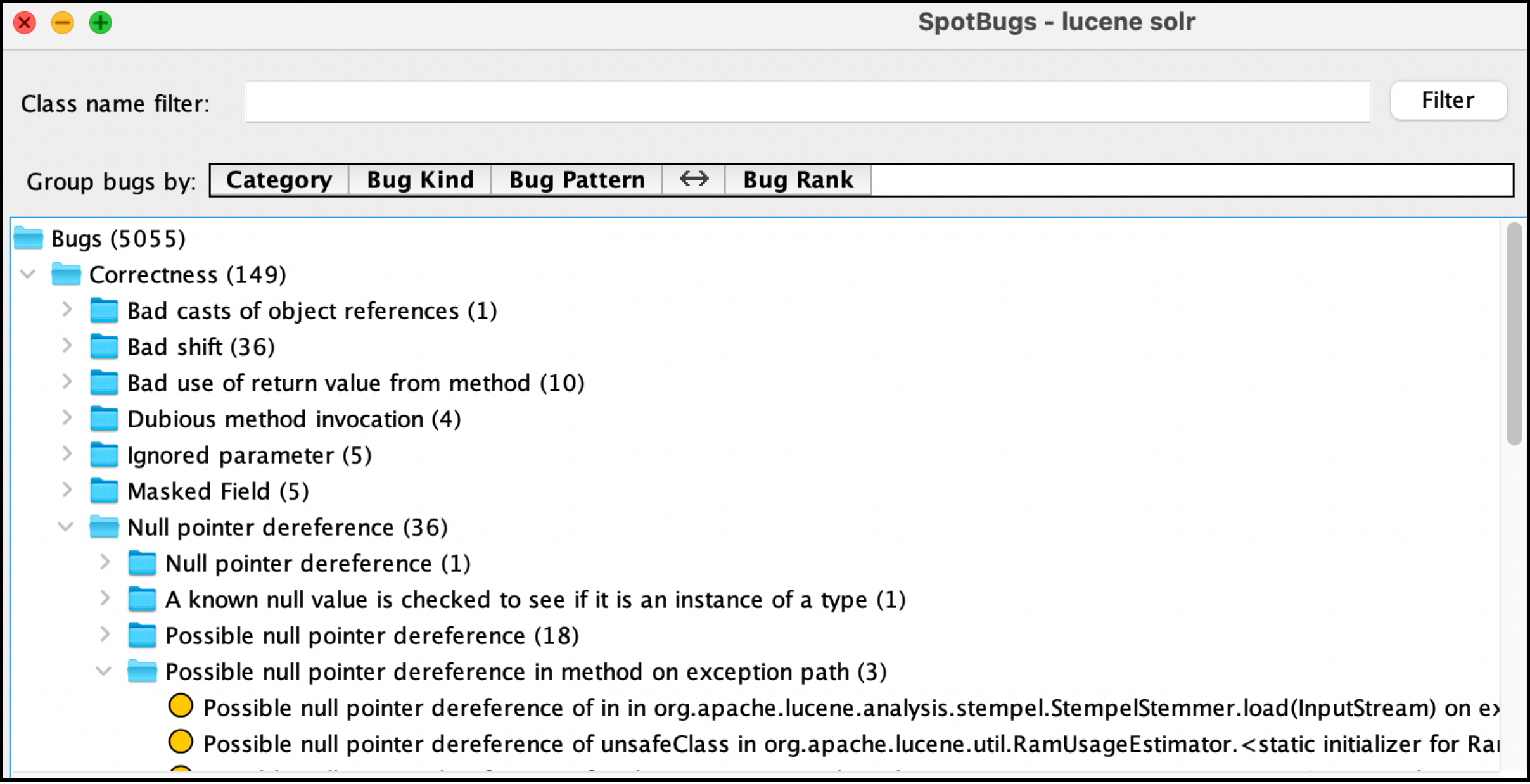}

  \caption{User interface of SpotBugs, a typical bug finding tool, which groups warnings by analysis kind or bug pattern. The lack of configurability for summarizing related warnings forces users to inspect warnings one-by-one.  }
  \label{fig:guis}
\end{figure}

% visualization
\indent{\textbf{\textit{Visualization.}}} With this focus on interactive sensemaking, \tool{}'s user interface is different from a typical user-interface of bug-finding tools such as SpotBugs (as shown in Figure~\ref{fig:guis}). Instead of displaying warnings in a tree view according to predefined filters or organizing them based on file locations, {\tool} organizes warnings with custom summaries created with interactive feedback.

% 1. evaluation: user study 
To evaluate \tool{}, we conducted a user study and a simulation experiment. First, we carried out a within-subject user study with 14 participants. The participants inspected null pointer dereference warnings reported by Infer and Spotbugs on two large mature projects (i.e., real warnings from real systems).
They were asked to articulate commonalities between uninteresting warnings.
We built a baseline for inspecting warnings one-by-one, similar to conventional tree-based organization bundled with Infer~\cite{inferexplore}, which reflects current developer practice~\cite{johnson2013don, christakis2016developers}. Rather than measuring how many warnings participants inspected, we evaluated how well they could articulate commonalities among related warnings.

A total of 55 different rules were formulated by 14 participants, with an average of 3.9 rules per participant.
When using \tool{}, the participants reported lower levels of mental demand (4.8 vs.~5.4 on a 7-point scale) and reported greater confidence (3.4 vs.~2.6 on a 5-point scale), compared to the baseline.
Using \tool{}, participants were able to better articulate commonalities among uninteresting warnings.
They also wrote longer, more detailed descriptions about the commonalities.
We found significant individuality among the summary rules constructed by individual participants, indicating that developers would desire greater configurability during their sensemaking process. Moreover, the significant individuality among rules confirms that warning inspection is an inherently subjective task, showing the benefit of personalized groups as opposed to universal pre-defined rules.

% 2. simulation
As the second evaluation method, we conducted simulation experiments to quantify the benefit of rule-based feedback and summarization. 
We observed, when a user provides rule-level feedback as opposed to instance-level feedback, rules inferred by \tool{} become aligned faster with interactive feedback. On average, for the warnings found in Alibaba Nacos, within 12 iterations, the rule alignment becomes over 80\%, and within 18 iterations, rule alignment becomes 100\%. 
%the rules were completely aligned. %Moreover at this point, the conciseness of the results improved by 1.79$\times$ when the rule-level feedback was used instead of iterating over the corpus of warnings one-by-one. 

In summary, this paper makes the following contributions:

\begin{enumerate}
    \item We design a personalized, interactive sensemaking workflow for tool-generated warnings, in which ILP-derived summary rules are iteratively refined through user feedback to highlight common characteristics by code location, type hierarchy, API signatures, or recurring expressions.
    \item In a within-subject user study, participants using \tool{} were better able to articulate commonalities among uninteresting warnings and reported greater confidence in their inspections.  
    \item A simulation study shows that summary rules align more quickly with rule-level feedback than with instance-level feedback.  
\end{enumerate}

The rest of this paper is organized as follows. 
Section~\ref{sec:usage} describes a usage scenario of \tool{}. 
Section \ref{sec:approach} introduces \tool{}'s active learning approach. 
Section~\ref{sec:design} presents the design of our user study. 
Section~\ref{sec:simulation} presents our simulation experiment design and its results.
Section~\ref{sec:discuss} discusses our study's implications and possible threats to validity.
Section~\ref{sec:related} presents related work.
Finally, we draw the conclusions of our work in Section~\ref{sec:conclusion}.

\section{Motivating Example}\label{sec:usage}
\lstset{
  language=java,
  basicstyle=\footnotesize\ttfamily,
  keywordstyle=\color{blue}\bfseries,
  commentstyle=\color{olive!60!black}\bfseries,
  stringstyle=\color{olive!60!black}\bfseries,
  escapechar=|,
  frame=single,
  breaklines=true,
  showstringspaces=false
}

\begin{figure*}[!h]
\centering

\begin{subfigure}[b]{.45\textwidth}
\begin{lstlisting}
Throwable ex = null;
int maxRetry = getProperty(ADDRESS_SERVER_RETRY_PROPERTY, Integer.class, DEFAULT_SERVER_RETRY_TIME);
for (int i = 0; i < maxRetry; i++) {
    try {
        ...
        success = true;
        break;
    } catch (Throwable e) {
        ex = e;
    }
}
if (!success) {
   |\colorbox{yellow}{throw new Exception$($SERVER\_ERROR, ex$)$;}|
}
\end{lstlisting}
\vspace{-1.5ex}
\caption{Infer warns that object \texttt{ex} can be null and is dereferenced by \texttt{new Exception(...)} at line 13.  This control-flow path is infeasible in practice if the property \texttt{ADDRESS\_SERVER\_RETRY\_PROPERTY} is non-zero. If \texttt{success} is false, \texttt{ex} is guaranteed to be not \texttt{null} and thus marked uninteresting by a user.}
\label{fig:case1}
\end{subfigure}
\hspace{0.02\textwidth}
\begin{subfigure}[b]{.45\textwidth}
\centering
\begin{lstlisting}
String server = getProperty(SERVER, StringUtils.EMPTY);
if (|\colorbox{yellow}{!server.startsWith(HTTPS\_PREFIX) \&\& }| |\colorbox{yellow}{!server.startsWith(HTTP\_PREFIX)}|) {    
    if (!InternetAddressUtil.containsPort(server)) {
            server = ... 
    }
    server = HTTP_PREFIX + server;
}
\end{lstlisting}
\vspace{-1.5ex}
\caption{Infer warns that \texttt{server} can be null and is dereferenced at line 2. This warning uses code expression, \texttt{getProperty}, similar to Figure \ref{fig:case1}, allowing a user to group similar warnings upon inspection of Figure \ref{fig:case1}}
\label{fig:case2}
\end{subfigure}

\vspace{0pt}

\begin{subfigure}[b]{.45\textwidth}
\centering
\begin{lstlisting}
String location = getLocation(LOGBACK_LOCATION);
try {
    ...
    LogbackConfigurator.configure(|\colorbox{yellow}{getResourceUrl(location)}|);
    ...
} ...
\end{lstlisting}
\vspace{-1.5ex}
\caption{Infer warns that \texttt{location} can be null and is dereferenced by \texttt{ResourceUtils.getResourceUrl} at line 4. This null pointer dereference risk is low, as \texttt{getLocation} returns a value configured within the application. This warning is a low-priority warning.}
\label{fig:case3}
\end{subfigure}
\hspace{0.02\textwidth}
\begin{subfigure}[b]{.45\textwidth}
\centering
\begin{lstlisting}
String readFile(String path, String fileName) {
    ...
    File file = openFile(path);
    if (|\colorbox{yellow}{file.exists()}|) {
        ...
}
\end{lstlisting}
\vspace{-1.5ex}
\caption{A warning representing a genuine null pointer dereference risk. A call to \texttt{openFile} can return null, making object \texttt{file} null and \texttt{file.exists} to reference a null value. Since this warning is interesting, \tool{} infers rules with predicates such that they do not match previously marked uninteresting ones, e.g., a \texttt{String} return type. } 
\label{fig:positive1}
\end{subfigure}%

\vspace{0pt}
\caption{Figures \ref{fig:case1}, \ref{fig:case2} and \ref{fig:case3} show three  uninteresting warnings. These null pointer dereferences marked in yellow are not worthwhile to address, since these  objects can be assumed to return valid values at runtime; With active feedback from a user on (a), and (b), \tool{} derives a summary rule to ignore uninteresting warnings in package \texttt{com.alibaba.nacos.client}. Then a user examines another warning (c) in the same package. When a user highlights \texttt{getProperty} in (b) to provide a hint, \tool{} refines the rule. }
\label{fig:motivating}
\end{figure*}

\noindent\textit{Overview.} To motivate our approach, we introduce a running example from Infer’s null-pointer analysis (Fig.~\ref{fig:motivating}). The yellow highlights mark the null pointer dereference warnings reported by the analyzer. Although these warnings look similar, their actionability differs: (a)–(c) are less interesting warnings, whereas (d) reflects a genuine defect. We use this example to first show why one-by-one inspection may hinder sensemaking, and then to demonstrate how \tool{} turns a few pieces of user feedback into summary rules that group similar uninteresting warnings.

\subsection{Lack of sensemaking support in existing tools} 
Suppose that Alice is given the task of inspecting null pointer dereference warnings generated by a static analyzer (Infer). 
Upon running it, Alice is overwhelmed by the large number of tool-generated warnings. %Suppose Alice browses the most common type of warnings: \texttt{Null Pointer Dereference}.
After inspecting several of them, Alice is frustrated and cannot easily make sense of which warnings should be acted upon. 
Figure \ref{fig:case1} shows an example null pointer dereference warning reported by Infer. 

\paragraph{\textbf{Difficulty of sensemaking when inspecting warnings one-by-one}}
Reviewing each warning one-by-one, Alice determines if each warning is worthwhile to address or should be ignored. 
After inspecting a few, Alice notices that several uninteresting warnings share similar characteristics, for example, being located in the same package or invoking similar API calls. 
This process remains burdensome and frustrating, as Alice cannot easily check and articulate the underlying characteristics of similar recurring warnings. 

\paragraph{\textbf{Difficulty of sensemaking when using pre-determined filters}}
Alice may decide to determine a set of rules to triage groups of related tool-generated warnings. 
She observes that many warnings from the package  \texttt{com.alibaba.nacos.persistence} are uninteresting for the same reason (e.g., a field set with the value fetched from a database is not \texttt{null}). 
At this point, she manually writes a script to identify warnings in the package \texttt{com.alibaba.nacos.persistence}.
Consequently, she observes that several warnings related to a field set through serialization tend to be less interesting to her; she hypothesizes that these warnings tend to be located in package \texttt{com.alibaba.\-nacos.\-config.server}. 
At the same time, she realizes that such warnings should be refined by subtyping relations, since they are located in classes implementing database-related interfaces. 
To mitigate this, she has to revise the existing script or write another one. 
Alice realizes that it is difficult to pre-define a set of rules, since the sensemaking process requires interactive navigation, search, and summarization of similar warnings. 

\subsection{\tool{}'s interactive sensemaking}
Alice then gives \tool{} a try. After labeling a few uninteresting warnings, she notices that the tool automatically proposes summary rules to group similar ones. These rules draw on containment relations, type hierarchies, API signatures of implicated methods, and recurring expressions. For instance, a rule may capture all warnings in the package \texttt{com.alibaba.nacos.config.server}. Below, we discuss two kinds of sensemaking processes in more detail.

\paragraph{\textbf{Example sensemaking with a containment relation.}}
Alice can view and provide feedback on multiple warnings simultaneously at the rule level. 
For instance, she notices that several warnings are clustered within the same package \texttt{com.alibaba.nacos.config.server}. 
Since these warnings all originate from the configuration server module, she realizes they are related to internal configuration handling and not to user-facing functionality. 
\tool{} proposes a rule based on this containment relation, grouping all warnings within the package together. 
After reviewing the matched warnings and confirming that none are of interest, Alice labels the entire group as \textit{``Uninteresting''}, expediting her inspection by resolving dozens of warnings in one action.

\paragraph{\textbf{Example sensemaking with salient code expressions.}} For the remaining warnings, Alice labels a few more and \tool{} proposes a refined rule grouping warnings in the package \texttt{com.alibaba.nac-} \texttt{os.client}. 
As these warnings share similarities (see Figure~\ref{fig:case3}), she notices that the function call \texttt{getProperty()} is a marker of uninteresting warnings. 
She highlights \texttt{getProperty} in the code view to signal to \tool{} that this expression should be included when reformulating the rule. 
\tool{} then refines the rule to cover all warnings in \texttt{com.alibaba.nacos.client} that call \texttt{getProperty()}, enabling her to quickly summarize and dismiss other similar warnings.

\section{Approach: Sensemaking with Interactive Review and Learning}\label{sec:approach}
\begin{figure}[t]
  \includegraphics[width=0.5\textwidth]{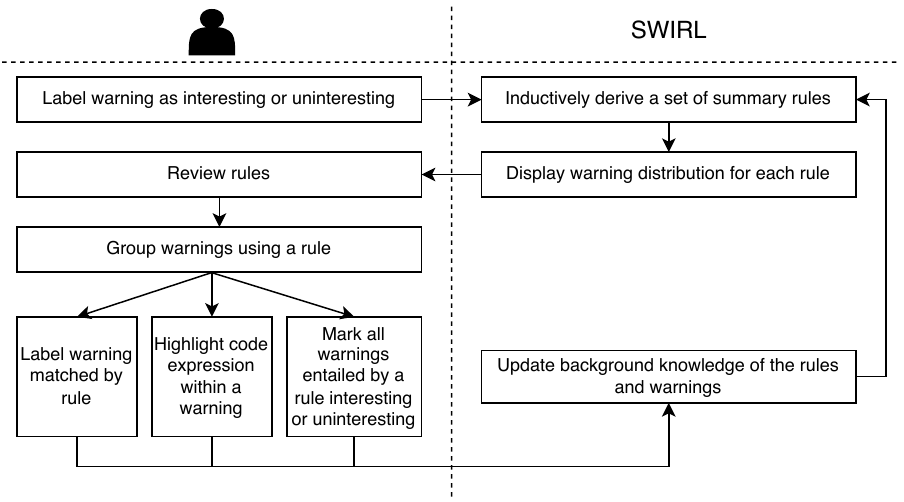}
\centering
\caption{\tool{}'s workflow: A user can provide two modes of interactive feedback: (1) instance-level feedback for each warning, (2) marking all warnings as interesting or uninteresting per rule, and (3) clicking and highlighting code expressions in a code snippet view to guide rule inference. Based on feedback, \tool{} updates its candidate rules and displays a distribution of matched and unmatched warning instances for each summary rule. 
% The user can then save and reuse derived rules for examining entailed warnings. 
}
  \label{fig:workflow}
\end{figure}

\begin{figure*}[!h]
  \centering
  \begin{subfigure}{0.48\textwidth}
    \centering
    \includegraphics[width=\linewidth]{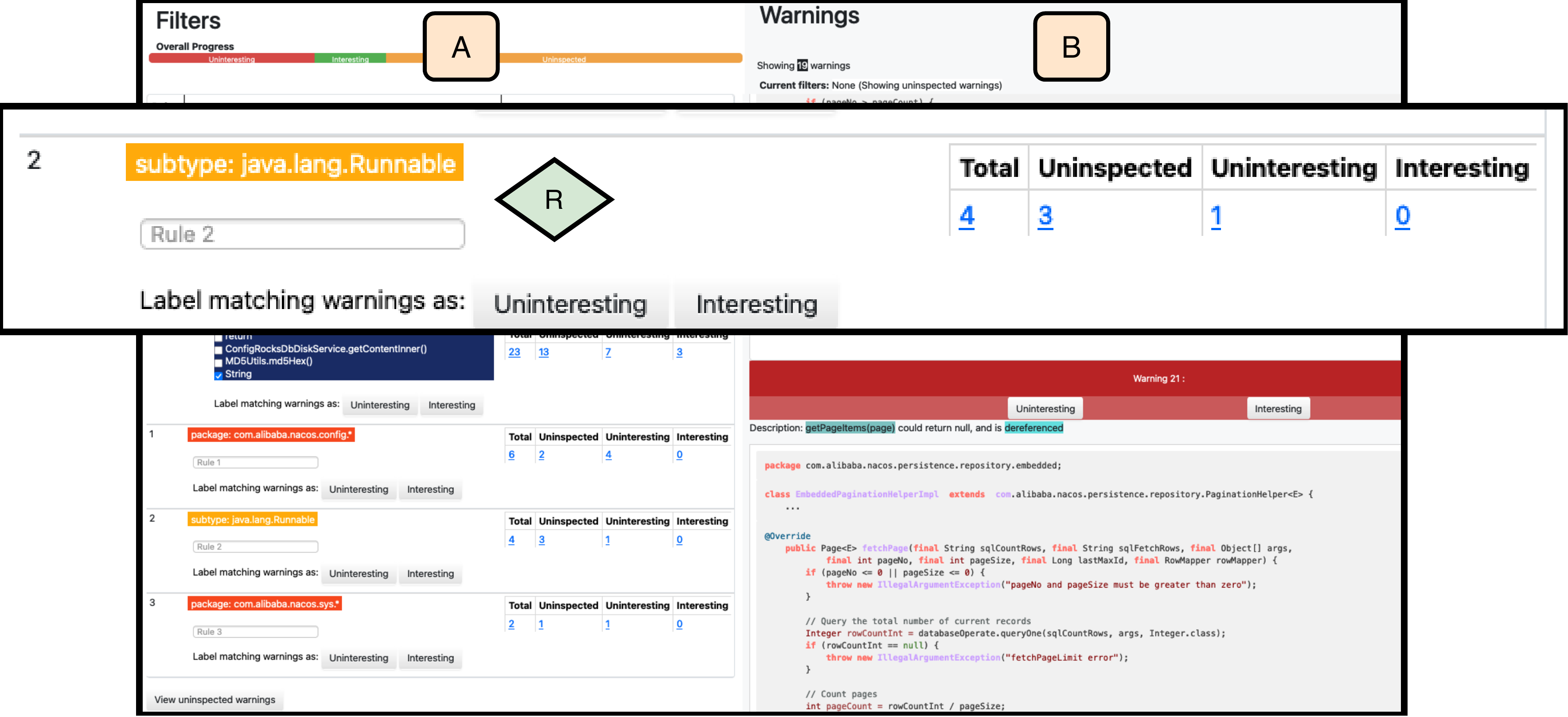}
    \caption{\tool{} has derived a candidate rule [see diamond R], which has a predicate checking for subtypes of \texttt{IntIterator},
    which detects warnings reported in classes with \texttt{Runnable} as a supertype. In the individual warning view [see square B], each warning is shown with its matching rule numbers, alerting users that the given warning might be similar to a previously inspected warning.}
    \label{fig:containment_rules}
  \end{subfigure}
  \hfill
  \begin{subfigure}{0.48\textwidth}
    \centering
    \includegraphics[width=\linewidth]{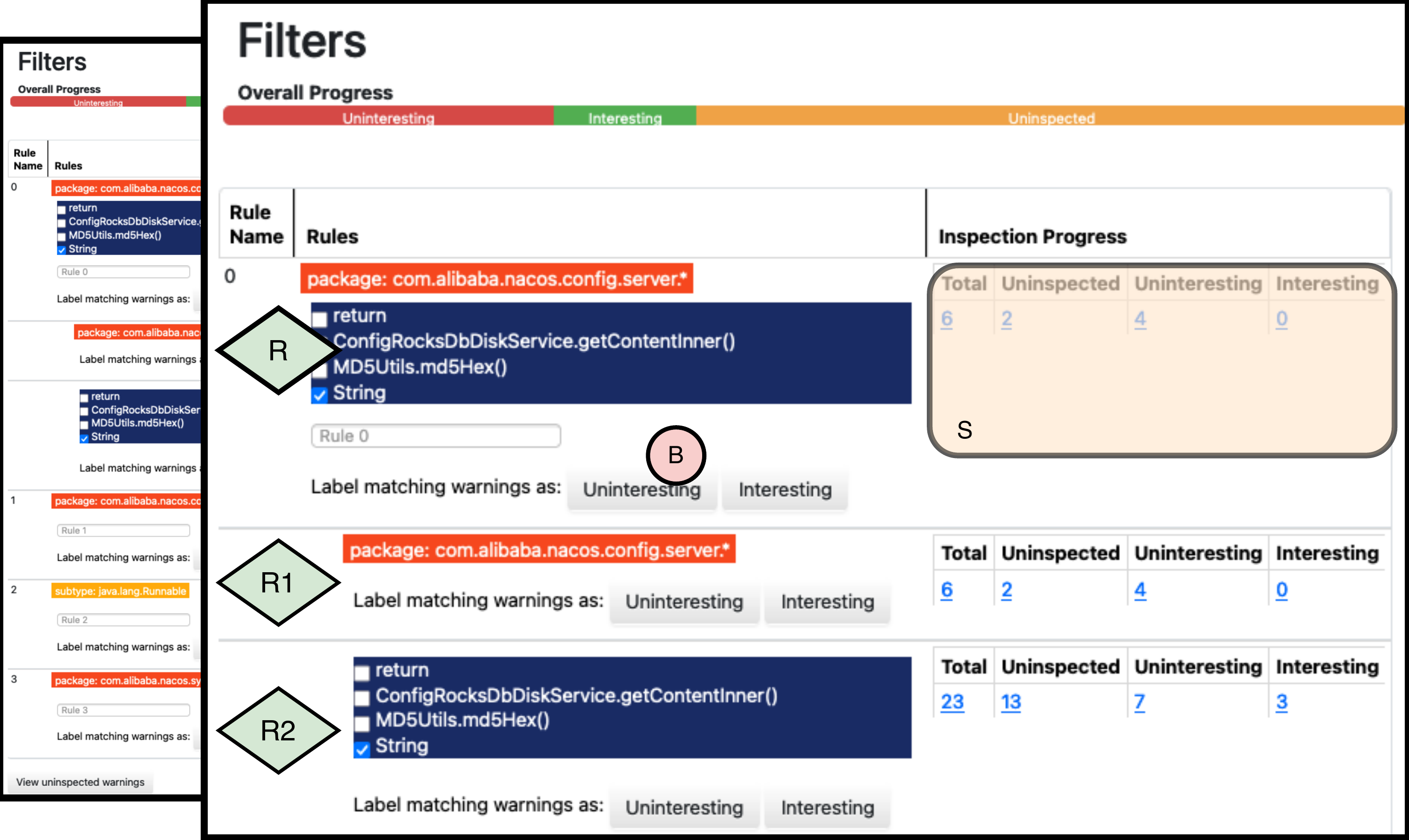}
    \caption{A rule [see diamond R] can combine both containment relations [see diamond R1] and code expressions [see diamond R2].
    Beside each rule, the statistics [see rectangle S] are shown: the total number of warnings, the number of uninspected warnings, the number of warnings already marked as uninteresting vs.~interesting. To expedite the sensemaking process, a user can provide feedback at the rule-level [see circle B].}
    \label{fig:conjunction}
  \end{subfigure}
  \caption{Containment and conjunction rules.}
  \label{fig:combined_rules}
\end{figure*}

This section introduces \tool{}’s interactive rule-learning workflow (Fig.~\ref{fig:workflow}), casting summary-rule derivation as an active-learning task solved via inductive logic programming.
%\miryung{the goal of this step is to do X becaused justification}
%\burak{Added an introduction sentence}
\tool{} eases the process of sensemaking tool-generated warnings by proposing summary rules that group related warnings. Rules are iteratively refined. 

\subsection{Active learning-enabled rule formulation}
We formulate the task of deriving summary rules as an \textbf{Active Learning} problem as follows:

\textbf{Inputs:} 
%\tool{} uses the following inputs:
\begin{enumerate}
    \item Labels of each inspected warning.
A warning, $w$, is labeled either as uninteresting ($w \in E^-$) or interesting ($w \in E^+$). 
    \item Background knowledge. A knowledge base with predicates on containment relationships and salient code expressions.
\end{enumerate}

\textbf{Output:}
\tool{} returns a set of summary rules, $R$. 
Each rule, $r \in R$, is a conjunction of predicates.
The goal is for each rule to group related warnings.
A warning matches a rule if every predicate in the rule is associated with it.

\textbf{Active Learning.} 
\tool{} has several requirements:
\begin{enumerate}
    \item To minimize the time users spend analyzing individual warnings one by one, \tool{} should provide help early in the inspection process.
    Thus, even when the total number of warnings labeled ``Uninteresting'' ($|E^-|$) and ``Interesting'' ($|E^+|$), is much lower than the total number of warnings ($|W|$), \tool{} should be able to derive rules.
    \item Rules derived from limited data are unlikely to accurately match the user's intent. 
    Thus, their refinement is necessary to formulate meaningful rules.
    We favor starting with a general set of rules, which maximizes the number of matching, uninspected warnings.
    \item Following Occam's Razor~\cite{blumer1987occam}, \tool{} should propose as few rules as required. 
    \item No warnings are removed; rather custom rules are dynamically created by {\tool} to group and highlight the common characteristics of similar warnings.
\end{enumerate}

\subsection{Inductive logic programming for deriving summary rules}

\begin{algorithm}[t]
\caption{Algorithm for deriving summary rules using Inductive Logic Programming~\cite{cropper2022inductive}, and updating the background knowledge based on the user's labeling and code highlighting feedback. }
\label{alg:ilp}
\begin{algorithmic}[1]
\Require $W \gets$ set of warnings under inspection
\State $KB \gets \text{extract\_containment\_facts(W)}$
\State $E^+ \gets \emptyset$
\State $E^- \gets \emptyset$

\While{user still has inspections}
    \If{user provided a new label}
        \State $w, label \gets $ \Call{fetchNewLabel()}{}
        \If{$label$ is Uninteresting}
            \State $E^- \gets E^- \cup w$
        \ElsIf{$label$ is Interesting}
            \State $E^+ \gets E^+ \cup w$
        \EndIf
    \EndIf
    
    \If{user selected code expression}
        \State $code\ expression \gets $ \Call{fetchCodeExpression()}{}
        \State $KB \gets$ \Call{updateKB}{$KB, W, code\ expression$}
    \EndIf
    
    \State $R \gets$ \Call{ILP\_Solve}{$E^+, E^-, W, KB$}
    \State \Call{PresentToUser}{$R$}
\EndWhile

\end{algorithmic}
\end{algorithm}

As shown in Algorithm \ref{alg:ilp}, \tool{} utilizes inductive logic programming~\cite{cropper2022inductive} to derive a set of summary rules.
\tool{} maintains a background knowledge base that is initialized by extracting facts of the implicated code's scope (package, class name), type hierarchy (subtyping/interfaces implemented) and other API signatures (return type, fields used) (line 1).
The user provides feedback by marking each warning instance as interesting or uninteresting or by selecting a salient code expression for the warning. If the user provides instance-level feedback, the set of uninteresting warnings and  interesting warnings are updated (lines 5--12). 
If the user highlights code expression as a hint in a code snippet window, the knowledge base is then updated by extracting new facts related to the code expression (lines 13--16). 
Note that we do not initialize the background knowledge with all possible code expressions, given the large space of possible rules, if every code expression is considered. Instead, \tool{} relies on the user to provide hints by highlighting salient code expressions.

To derive a set of rules,
\tool{} encodes the knowledge base as a logic program.
The rules group potentially related warnings for user inspection.
In our problem domain, a hypothesis is an assignment of predicates to rules such that the rule should not match the interesting examples ($E^+$), while matching as many uninteresting examples ($E^-$) (line 17).
Ties between possible hypotheses are broken by the number of matched warnings, following the objective of maximizing the rules' generality.
We use an off-the-shelf logic program solver~\cite{gebser2010gringo} to search for candidate hypotheses. 
% Finally, the inferred hypothesis is presented to the user as the proposed set of rules (line 18).
As shown in Figure \ref{fig:containment_rules} (R), \tool{} displays the rules, indicating the number of warnings matched by each  rule.

%\begin{figure}[t]
%  \includegraphics[width=0.5\textwidth]{images/tutorial/containment_rule.pdf}
%\centering
%\caption{\tool{} has derived the rule (R), which has a predicate checking for subtypes of \texttt{IntIterator}, 
%which detects warnings reported in classes with \texttt{Runnable} as a supertype. In the individual warning view (B), each warning is shown with its matching rule numbers, alerting users that the given warning might be similar to a previously inspected warning.}
%  \label{fig:containment_rules}
%\end{figure}

\subsection{Rule design}
%\miryung{add related work and justication on why we chose containment, salient expression and composition}
%\burak{Changed "Containment relationships" and "Composite rules" paragraphs.}
\textbf{Containment relationships.}
Prior work indicates that static-analysis alarms often cluster containment relationships (such as package, class, and type hierarchy)~\cite{kremenek2004correlation, buckers2017uav}. \tool{} infers rules as a composite of predicates with the following structural relations.

\begin{enumerate}
    \item Package: The package where the warning was reported (\texttt{package(X)})
    \item Classname: The name of the class where the warning was reported (\texttt{classname(X)})
    \item Return type: The return type of the method where the warning was reported (\texttt{rettype(X)})
    \item Fields used: The fields of the class used in the method (\texttt{fields(X)})
    \item Subtyping: The parent classes and interfaces of the class (\texttt{subtype(X)})
\end{enumerate}

Fry and Weimer report that static analysis reports in the same file and concerning the same program element (e.g. a particular function or variable) are likely duplicates that can be handled together~\cite{fry2013}. Such evidence supports our decision to use containment predicates (package name, class name, parent type) as they capture a natural facet for grouping related warnings in the same context.

\textbf{Salient code expressions.}
A user can select expressions to provide hints with code features associated with warnings, which can be specific method calls~\cite{van2020tailoring} or API usage conventions~\cite{uddin2011analyzing}. For a code expression highlighted by a user, \tool{} extracts the methods called (e.g., \texttt{getProperty()}), the types of the variables (e.g., \texttt{String}), the values of literals (e.g., \texttt{os.name}), control-flow elements, such as \texttt{if-then} statements, as well as conditionals such as a null-check. \tool{} encodes each code expression as a predicate, adding them to the background knowledge for rule inference. 
This grouping using salient expressions is motivated by prior findings highlighting the importance of capturing API usages through common method calls or similar expressions ~\cite{zhang2018code,nguyen2010libsync}. 
% \sout{Hence we enable users to highlight such expressions as cues for refining rules.}

%\begin{figure}[t]
%  \includegraphics[width=0.5\textwidth]{images/tutorial/conjunction_rules.pdf}
%\centering
%\caption{A rule (R) can combine both containment relations (R1) and code expressions (R2). 
%Beside each rule, the statistics (S) are shown:  the total number of warnings, the number of uninspected warnings, the number of warnings already marked as uninteresting vs.~interesting. If a user believes a rule entails only uninteresting warnings, they can mark all matching ones as uninteresting (B).}
%  \label{fig:conjunction}
%\end{figure}

\textbf{Composite rules.}
\tool{} can derive rules that combine containment relationships with specific code expressions. 
For example, it can generate a rule to detect invocations of a given method ``\texttt{getProperty()}'' within the package 
\texttt{com.alibaba.nacos.core.cluster}, formally expressed as
\texttt{package(com.alibaba.nacos.core.clu\-ster)} and \texttt{codeElement(getProperty)}.

Conjoining containment predicates with salient code expressions yields more precise, developer-aligned groupings than any single facet; prior work shows that combining structural and code-level signals improves triage and reduces inspection effort (e.g., \cite{fry2013, liargkovas2023quietingstaticstudystatic}).
% \sout{Moreover, other prior work on multidimensional concerns found that similar code snippets can be grouped along multiple concerns~\cite{ossher2002,menzies2000validating,kersten2006}, justifying the use of composite rules that can combine multiple common characteristics.
% Thus we adopt interpretable containment predicates (package, class, return type, fields, subtyping) together with salient code expressions as the building blocks for summary rules (e.g., \cite{van2020tailoring, uddin2011analyzing}).}

\subsection{Distribution of matched inspected warnings}
\textbf{Sensemaking statistics.}
In \tool{} UI, the column ``Inspection Progress'' summarizes the distribution of warnings for each rule, as shown in Figure \ref{fig:conjunction} [rectangle S] to indicate the sensemaking progress with formulated rules and to guide users in assessing if a candidate rule meaningfully groups related warnings. By clicking on the underlined numbers, a user can jump to the corresponding warnings matched by the rule.
%and zoom in their inspection. 

\textbf{Applying a rule to label all matching warnings.}
If a user believes that a rule accurately describes a group of similar uninteresting (or interesting) warnings, they can click the button beside ``Label matching warnings as uninteresting (or interesting)'' as shown in Figure \ref{fig:conjunction} [see circle B]. A user can also rename a numbered rule with a more descriptive name to create meaningful abstraction for grouping related warnings.

\section{User Study}\label{sec:design}
\noindent Our user study answers the following research questions:
\begin{enumerate}
    \item RQ1. How much variation exists among the summary rules interactively formulated by different users?
    \item RQ2. How does \tool{} influence developers' ability to identify commonalities between warnings, their cognitive load and their confidence in their inspections?
    \item RQ3. What features of \tool{} are useful for formulating rule abstraction during the sensemaking process?
\end{enumerate}

% \begin{table}[t]
%   \caption{Subject programs used in the user study. The warnings generated by Infer and SpotBugs from the null pointer dereference analysis are used in the study. \hj{I added this table, but now I think we might want to remove this since it doesn't offer much useful information }}
%     \label{tab:datasets}
%     \begin{tabularx}{\linewidth}{ X   l r  }
%       \textbf{Project}  &  
%       \textbf{Static analysis } & \textbf{\#warnings }  \\ 
%       &  & (\% false positives) \\
%       \hline
%       alibaba/nacos  & Infer & 58 (43\%)\\
%       apache/lucene-solr  & SpotBugs  & 84 (56\%) \\
%     % \bottomrule
%   \end{tabularx}
% \end{table}

The goal of the user study is to assess if \tool{} enables participants to better articulate common characteristics of warnings that can inform downstream action (e.g., de-prioritization, fixing the warnings, or policy updates).
% \sout{\tool{} emphasizes \textit{grouping}, \textbf{not} filtering.
% \tool{} does not remove any warnings but rather formulates custom summary based on user feedback.}
Accordingly, we measure the quality of groupings by assessing whether the participants could articulate common characteristics of related warnings, comparing \tool{}’s effect on sensemaking against a baseline approach with inspecting warnings individually.

%\burak{Added this introductory paragraph to emphasize the goal of the user study.}

\textbf{Baseline.}
To reflect how developers interact with static analysis tools in practice~\cite{johnson2013don, christakis2016developers}, we constructed a downgraded version of \tool{}, where a user can browse warnings only one-by-one, similar to a typical interface of a warning list view bundled with Infer~\cite{inferexplore}. As shown in Figure~\ref{fig:guis}, SpotBugs display warnings one by one for each analysis kind. 

%\miryung{the goal of evaluation is to triage and articulate common characteristics of warnings to act on. emphasize that none was removed.}
\textbf{Participants.}
We ran a counter-balanced within-subject crossover study with 14 participants: 10 Ph.D. students, 2 master’s students, and 2 industry developers, recruited from our CS department and professional networks. Experience ranged from 1–3 years (2), 4–6 years (3), 7–10 years (7), to $>$ 10 years (2). Exposing every participant to both tools in randomized order controls individual variability and enables paired comparisons with a modest sample size~\cite{Vegas_Apa_Juristo_2016, chasins2021pl}. Such within-subject studies with 8–16 participants are standard practice across Software Engineering~\cite{Arteaga_Garcia_Nicolaci_Pimentel_Feng_Gerosa_Steinmacher_Sarma_2024, Kang_Wang_Kim_2024, Ganji_Alimadadi_Tip_2023}, HCI~\cite{Horvath_Myers_Macvean_Rahman_2022, Suh_Min_Palani_Xia_2023, Huh_Pavel_2024}, and sensemaking research~\cite{selenite, synergi, beyond_code_gen, dreamsheets, confidence_highlighting}.

\textbf{Protocol.}
We conducted a 1-hour user study with each participant, involving both \tool{} and the baseline tool. 
The tool order and assigned tasks (Task A or Task B) were counter-balanced randomly. 
The study began with a 3-minute pre-study survey about the participants' background, followed by a 5-minute tutorial.
Participants then completed 10 minutes of warm-up questions, inspecting a few null pointer dereference warnings from Infer.
They proceeded to the main tasks, each lasting 10 minutes, where they inspected up to 25 warnings using the assigned tool.
%Participants completed both case studies, with each task set for 10 minutes but with the option to finish early. 
This time limit reflects typical usage of static analysis tools, 
where developers are reluctant to spend a significant amount of time analyzing the warnings, which we further verified in our pilot study.
It was also {\em not} required for a user to review all warnings. 
After each task, participants filled out the NASA--TLX and post-task surveys (5 mins each). 
Finally, they completed a 12-minute post-study survey rating their experience with both tools and the usefulness of \tool{}'s UI features.

\noindent\textbf{Tasks.}
The user study consists of two tasks. For \textbf{Task A}, users inspected the warnings generated by Infer on Alibaba Nacos~\cite{alibaba/nacos_2024}. 
For \textbf{Task B}, users inspected warnings generated by SpotBugs on Apache Lucene-Solr~\cite{apache/lucene-solr_2024}. These subject programs were used by prior work
~\cite{kharkar2022learning, wang2018there, yang2021understanding, kang2021active, yedida2023find}.

% \begin{figure*}[]
%     \centering
%     \includegraphics[width=\textwidth]{images/boxplot_combined_superlarge_means.pdf}
%     \caption{Assessments of cognitive burden for \tool{} and the baseline by the participants. \tool{} reduced cognitive load for participants by decreasing the level of mental demand they experienced (5.4 to 4.8), slowing the perceived pace of the task (5.2 to 4.7), and lowering the frustration level (4.8 to 3.9). \hj{When we have an issue with lack of space eventually, we might want to remove this figure?}}\burak{Yes, we can remove this.}
%     \label{fig:post_task}
% \end{figure*}

\subsection{RQ1. Variation in user-formulated rules}

\textbf{Number and variance of derived rules.} In total, 55 different rules were derived and used by the participants. 
Each participant formulated an average of 3.9 rules, and created at least one unique rule.
% \sout{This highlights that during sensemaking, users develop \textit{personalized interpretations} of warnings, focusing on aspects relevant to their individual perspectives.}
This suggests that users develop \textit{personalized interpretations} of warnings as they make sense of them,
% The high degree of variations among the rules confirms
and that developers desire configurability.
% and 
As such,
a one-size-fits-all policy with universal and pre-defined groups may not work for summarizing warnings. 

\begin{table}[!h]
\centering
\footnotesize
\begin{tabular}{p{6.5cm}|p{1.5cm}}
\textbf{Rules} & \textbf{Participants} \\ \hline
Warnings with code expression \texttt{Collection\-Utils.isNotEmpty()} & P1, P4, P5, P11 \\ 
Warnings with return type \texttt{boolean} matching \texttt{Iterator.*} & P2 \\ 
Warnings in package \texttt{org.apache.solr.*} with return type \texttt{void} & P2 \\ 
Warnings matching statement \texttt{File.is*} & P3, P6 \\ 
Warnings with return type \texttt{Page\textless{}E\textgreater{}} for null object & P12 \\ 
Warnings involving interface \texttt{NamedListInit\-ializedPlugin} & P13 \\ 
\end{tabular}
\caption{Example summary rules constructed by participants.}
\label{tab:rules}
\end{table}

Table \ref{tab:rules} shows examples of participant-formulated rules guided by \tool{}'s active learning. 
A complete table is provided in our replication package~\cite{allRules}. The rules formulated by the participants capture a wide range of characteristics, including code expressions for third-party libraries (e.g., \texttt{Collection\-Utils}), file handling (e.g., \texttt{File\-.list\-Files()}), and specific API usage (e.g., use of \texttt{Iterators}). Other rules express containment relations in terms of the package name (e.g. \texttt{org.\-apache.\-solr.*}) or subtyping relations (e.g., \texttt{Named\-List\-Initialized\-Plugin}).
We observed utilization of salient code expressions for grouping, i.e., 4 out of 14 participants formulated a summary rule with code expression hint \texttt{Collection\-Utils\-.is\-Not\-Empty()}.

\noindent\textbf{Types of rules.}
Participants used rules with different kinds, 51\% of the derived rules were based on containment relationships (package name, class name, return type, used fields, and subtyping), and 40\% were based on code expression hints highlighted by participants. The other 9\% of the rules were composite rules consisting of both containment relationships and code expressions.

\vspace{-1mm}
\begin{tcolorbox}[colback=black!5!white,colframe=white!50!black]
\textbf{RQ1.} Of the 55 rules created, each participant contributed at least one unique rule, demonstrating individual approaches to sensemaking tool-generated warnings.
\end{tcolorbox}
\vspace{-2mm}

\subsection{RQ2. Impact on sensemaking}
\noindent\textbf{Identifying the commonalities between uninteresting warnings.}
% The participants were instructed to identify the root causes of the false positives.
Participants were instructed to articulate possible root causes/commonalities of the related warnings after each task. We found that participants could list 13\% more common symptoms of related warnings (1.53 vs 1.35), when using \tool{} compared to the baseline on average. Participants also provided more detailed descriptions, writing an average of 4.6 sentences compared to just 3.1 sentences when using \tool{} vs.~the baseline, indicating that they were able to better recognize similarities between related warnings. Participants using \tool{} pointed out specific code elements in their articulation more frequently than the baseline. For example, when using \tool{}, P6 pointed out that \texttt{isDirectory} acted as a guard for \texttt{listFiles}, making its null pointer dereference risk a low priority. 
P11 observed that some functions from third-party libraries can be assumed to handle null inputs gracefully (e.g., \texttt{Collections.isNotEmpty}). 
% \smallskip

\begin{figure}
    \centering
    \includegraphics[width=\columnwidth]{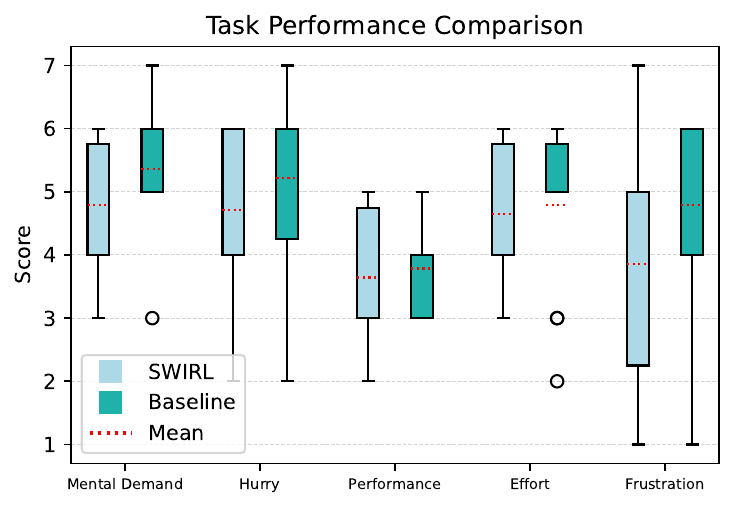}
    \caption{Assessments of cognitive burden for \tool{} and the baseline by the participants. \tool{} reduced cognitive load for participants by decreasing the level of mental demand they experienced (5.4 to 4.8), slowing the perceived pace of the task (5.2 to 4.7), and lowering the frustration level (4.8 to 3.9).}
    \label{fig:boxplot_nasa}
  \end{figure}

\noindent\textbf{Cognitive load.}
Results show that \tool{} was associated with significantly lower perceived \textit{mental demand} compared to the baseline ($-0.71$ points on the questionnaire scale, $p=0.043$). For \textit{hurriedness}, participants also tended to report lower values with \tool{} ($-0.65$ points), though this difference did not reach significance ($p=0.059$). No significant tool effects were observed for \textit{effort}, \textit{success}, or \textit{frustration}. Descriptive means mirror these patterns in Figure~\ref{fig:boxplot_nasa}: mental demand decreased (5.4$\rightarrow$4.8), the perceived pace slowed (5.2$\rightarrow$4.7), and frustration was lower (4.8$\rightarrow$3.9). Table~\ref{tab:boxplot_ease_of_use_confidence} further shows higher confidence with \tool{} (3.4 vs.~2.6) alongside a slight decrease in perceived ease of use (3.4 vs.~3.7). Taken together, these results suggest that \tool{} helped reduce participants' cognitive load when inspecting warnings, particularly in terms of mental demand. % in their inspections.  
% Qualitative coding further supports this interpretation: participants using \tool{} articulated clearer hypotheses about uninteresting warnings and reported higher confidence.

\vspace{-1mm}
\begin{tcolorbox}[colback=black!5!white,colframe=white!50!black]
\textbf{RQ2.} Participants using \tool{} better articulated commonalities among uninteresting warnings, listing 13\% more symptoms and writing longer, more detailed descriptions. In doing so, they also experienced lower cognitive load: mental demand was $\approx$0.7 points lower than the baseline $(p=0.043)$, and confidence was higher (3.4 vs.~2.6, cf.\ Tab.~\ref{tab:boxplot_ease_of_use_confidence}). 
\end{tcolorbox}
\vspace{-2mm}

% \begin{figure*}[!t]
%   \centering
%   \begin{subfigure}{\textwidth}
%     \centering
%     \includegraphics[width=\linewidth]{images/boxplot_combined_superlarge_means.pdf}
%     \caption{Assessments of cognitive burden for \tool{} and the baseline by the participants. \tool{} reduced cognitive load for participants by decreasing the level of mental demand they experienced (5.4 to 4.8), slowing the perceived pace of the task (5.2 to 4.7), and lowering the frustration level (4.8 to 3.9).}
%     \label{fig:boxplot_nasa}
%   \end{subfigure}
%   \hfill
%   \begin{subfigure}{0.5\textwidth}
%     \centering
%     \includegraphics[width=\linewidth]{images/plot_combined_superlarge.pdf}
%     \caption{Using \tool{}, the participants reported higher confidence  (3.4 vs.~2.6), with a slightly reduced ease of use (3.4 vs.~3.7).}
%     \label{fig:boxplot_ease_of_use_confidence}
%   \end{subfigure}
%   \caption{Plots for NASA--TLX Cognitive Load metrics and Confidence \& Ease of Use scores.}
%   \label{fig:aaabbb}
% \end{figure*}

%\begin{figure}
%\centering
%\includegraphics[width=\linewidth]{images/plot_combined_superlarge.pdf}
%\caption{Using \tool{}, the participants reported higher confidence  (3.4 vs.~2.6), with a slightly reduced ease of use (3.4 vs.~3.7).}
%\label{fig:boxplot_ease_of_use_confidence}
%\end{figure}

\begin{table}[h!]
\centering
\resizebox{0.8\columnwidth}{!}{
\begin{tabular}{l|cc}
\hline
 & \textbf{\tool{}} & Baseline \\
\hline
Confidence (1--5) & \textbf{3.43} & 2.64 \\
Ease of Use (1--5) & 3.43 & \textbf{3.69} \\
\hline
\end{tabular}
}
\caption{Using \tool{}, the participants reported higher confidence (3.4 vs.~2.6), with a slightly reduced ease of use (3.4 vs.~3.7).}
\label{tab:boxplot_ease_of_use_confidence}
\end{table}

%\burak{Updated the text about the statistical test.}
%\hj{we should say that because participants now had to think about the summary rules, there is additional cognitive burden, but nevertheless, participants were now more confident.}\burak{Done}

% \miryung{check this last sentence}

%Using \tool{}, participants spent part of their time trying to refine rules even when the rules did not help to find more uninteresting warnings but were able to come up with more possible commonalities between them.
% Nevertheless, despite the decrease in inspected warnings, the reusable rules would be useful as a filter for warnings reported in the future. 

% Graph: Approach, Task A , Task B
% showing # of suppressions
% % showing % correct suppressions
% \begin{figure}[t]
%   \includegraphics[width=\textwidth]{}
% \centering
% \caption{RQ2: }
%   \label{fig:impact}
% \end{figure}

\subsection{RQ3. User perception of the interface features}
% figures showing likert scale 
%\begin{figure}[t!]
%    \centering
%    \includegraphics[width=0.5\columnwidth]{images/plot_combined_superlarge.pdf}
%    \caption{RQ3: Using \tool{}, the participants reported higher confidence  (3.4 vs.~2.6), with a slightly reduced ease of use (3.4 vs.~3.7).}
%    \label{fig:confidence}
%\end{figure}

\noindent\textbf{Ease of use.} Despite \tool{}'s learning curve, participants found both the baseline and \tool{} easy to use (3.7 vs 3.5 on a 5-point Likert scale). 
%shown in Figure~\ref{fig:confidence}.
% which is expected since there is a significant amount of features introduced in \tool{}, which the users had limited time to learn in the scope of the 1-hour they had to complete the study (including filling the surveys and the completing the task with baseline). 
P2 stated: ``\textit{I think if I got the hang of \tool{} and understood the filters a little bit better, going through these errors and marking them would be much easier 
...}''
% \burak{11/14}
% than just doing it by hand. 
% ... }
% The filters and on-the-fly grouping were useful, ...} %but I don't think I utilized them in a way that really showcases the full utility of the tool...}''

% In comparison, the baseline is just consisting of a list of warnings with just the labeling buttons. Considering this, the similar scores of ease of use of the tools should be interpreted in favor of \tool{}. The distribution of responses is shown in Figure~\ref{fig:likert}.

\noindent\textbf{Containment relationships.} 10 out of 14 participants rated the feature of inferring rules with containment relations (package name, class name, used fields, return type, and subtyping) useful with a score of 3.2 out of 5. P3 wrote: ``\textit{Due to its similarity of those grouped alarms, I can understand the common root cause and skip the `warm-up' phase in my decision-making.}''
% \hj{@Burak: for each of the features, can we report the number of participants who gave a score >= 3? e.g., "10 out of 14 found X useful"}
% \burak{10/14}
% ``\textit{Because they could help me group them effectively in a readable way. Due to its similarity of those grouped alarms, I can understand the common root cause and skip the "warmup" phase in my decision-making.}''
%\hj{at the end of this subsection, we can also report this mix of results positively as an aggregate, e.g. "X of 14 participants found at least one feature of \tool{} useful"}

\noindent\textbf{Highlighting expressions as hints.}
12 out of 14 participants perceived highlighting code expressions to be more useful (a score of 3.7 out of 5) as it produced groups of similar warnings. 
P4 wrote that this allowed them to ``\textit{easily check all similar warnings at once"} since ``\textit{they might have the same reasons behind.}'' 
% \burak{12/14} 
% , so we speed up a lot when we check the warnings.}'' 

%\burak{This quote reads a little bit weird}\hj{I've split it into two quotes instead of just one.}
% ``\textit{I find it very helpful. Because by selecting them, we can easily check all the similar codes at one time. They might have the same reasons behind, so we speed up a lot when we check the warnings.}''

\noindent\textbf{Warning distribution.} \tool{} displays the number of matched warnings for each rule, including the proportion of uninteresting warnings already labeled by the user. 11 out of 14 participants found this feature useful (a score of 3.8 out of 5), as it allowed them to determine ``\textit{which rules are more general and are worth looking at.}'' 
% \burak{11/14}
% P3 said, ``\textit{It was very helpful to be able to view the un-inspected warnings that matched the pattern so that I could quickly go through them and just verify that it was the same function or the same guard.}% that the static analysis tool missed.}''
% This suggests that participants considered these statistics to determine if the rules were meaningful.
%However, this particular participant found that some of the statements he selected only matched the warnings they had already selected, so they found the feature to be useful in most cases.
% However, we also observed that participants tended to specialize a rule beyond what would be useful for the user study. 

\noindent\textbf{Label all warnings matched by a rule.}
Users can apply the same inspection decision to all warnings matched by the rule. 
8 out of 14 participants found this feature useful (avg. 3.2/5), particularly when they were confident the rule captured only warnings sharing the same inspection decision.
P9 said, ``\textit{Essentially the warnings are largely repetitive, as the static analyzer makes the same correct/mistaken decisions. So once the programmer has learned its pattern, it can be used to label all similar warnings.}''
% \burak{8/14}

% \textbf{Limitations and suggestions.}
% Participants pointed out \hj{??? we have a question about suggestions, so let's fill this in using a common suggestion. Not super important so let's do this last}
In aggregation, 13 out of 14 participants found \tool{} useful.
Moreover, they indicated that they would like to use \tool{} in the future (4.1 on a 5-point Likert scale).
% suggesting that \tool{} aids in reviewing tool-generated warnings.
% \burak{12/14}

\vspace{-1mm}
\begin{tcolorbox}[colback=black!5!white,colframe=white!50!black]
% The participants perceived \tool{} to be easy to use. 
\textbf{RQ3.} Participants found both containment relations and code expressions useful for constructing summary rules; they found being able to highlight concrete code expressions as hints useful for formulating custom abstraction.  

\end{tcolorbox}
\vspace{-2mm}

\section{Simulation}\label{sec:simulation}

%\begin{table}[!ht]
%\centering
%\resizebox{\columnwidth}{!}{%
%\begin{tabular}{|l|l|l|c|r|}
%\hline
%\textbf{Program}       & \textbf{Tool} & \textbf{Warning Type}       & \textbf{\# Warnings} & \textbf{LOC} \\ \hline
%Apache Lucene          & SpotBugs      & Null Pointer Dereference    & 25                  & 1,308,528   \\ \hline
%Alibaba Nacos          & Infer         & Null Pointer Dereference    & 25                  & 320,761     \\ \hline
%PrestoDB Presto        & Infer         & Resource Leaks              & 34                  & 1,693,230   \\ \hline
%Apache Dubbo           & CodeQL        & Taint Flows                 & 91                  & 399,576     \\ \hline
%\end{tabular}%
%}
%\caption{The subject programs, the tools, and warning types investigated in our simulation study.}
%label{tab:subject_programs}
%\end{table}

To complement and further verify the user study findings, we conducted a simulation study. While the user study demonstrated that \tool{} improves sensemaking quality and reduces cognitive load, it could not isolate the specific benefit of rule-level feedback over instance-level feedback at scale. The simulation addresses this by assessing the impact of inspection-order heuristics and feedback granularity (rule vs.\ instance) on the alignment of learned rules with user feedback, evaluated on the same data we used for \textbf{Task A}.
% Based on the descriptions of what constituted an unactionable warning from prior work~\cite{kharkar2022learning,van2020tailoring,kang2022detecting},
% we obtained the ground truth labels by inspecting and by manually validating the warnings. \hj{I think we need a different word other than "ground truth", otherwise it sounds like there is an absolute notion of "false positive" and we fall into the same hole as the previous submission. The reviewers may ask us if the users in the user study were "accurate". }

\subsection{Simulated Behavior}
In the user study, we observed that participants employ three heuristics to examine tool-generated warnings: 
\begin{itemize}
    \item \textit{Heuristic 1.}~Inspect \textbf {shorter} warnings first. Observed among participants [$P1$, $P2$, $P4$, $P5$, $P8$, $P9$, $P10$].
    \item \textit{Heuristic 2.}~Inspect warnings in terms of \textbf{similar API calls} first. Observed among participants [$P1$, $P2$, $P3$, $P5$, $P7$, $P8$, $P9$, $P10$].
    \item \textit{Heuristic 3.}~Inspect warnings located with \textbf{similar container names} (e.g., the same package name, the same class name, or the same file prefix). Observed among participants [$P2$, $P5$, $P9$].
\end{itemize}
% \miryung{Burak, can you add corresponding participant numbers that support each  heuristic?}
% \burak{Added - there are also people responded to the surveys that they used some heursitics. Should I be mentioning these?}

%Heuristic 1 prioritizes the shortest warning for inspection. 
%Heuristic 2 prioritizes uninspected warnings that share at least one API call with an inspected warning. 
%Heuristic 3 prioritizes warnings that share the same package 
% (i.e., \texttt{package.com.\-alibaba.nacos.\-sys.utils} and \texttt{package.com.alibaba.\-nacos.client.\-auth.ram.utils}) 
%with an inspected warning. 
%Our simulated user combines Heuristics 1, 2, and 3 by randomly using one heuristic by selecting an uninspected warning for inspection. 

% Based on the sorting order determined by the selected heuristic, 
In each interaction, the simulated user examines either one uninspected warning or an inferred rule. 
%We separate the warnings into two buckets: \texttt{to-be-examined} vs. \texttt{to-be-skipped}.  If the warning is in the \texttt{to-be-examined} bucket, a user proceeds with the next step.  This simulates the user's choice to skip inspection of this warning. 
With a probability $p$, a user provides feedback at the rule level. 
For example, $p=0.3$ means that a user provides rule-level feedback 30\% of interactions, i.e., marking all warnings entailed by the rule as uninteresting.
For providing feedback, we simulate a user making a decision for each of the selected warnings.
All heuristics means using combined three heuristics in tandem. In other words, our simulated user combines Heuristics 1, 2, and 3 by randomly using one heuristic by selecting an uninspected warning for inspection.

\subsection{Metric}
% We perform simulation experiments for four subject programs.
% In these simulations, we vary the heuristic used for ordering the inspection of the warnings (Heuristic 1, 2, 3, and 4), 
% and vary the preference of the user in providing rule-level feedback, changing this probability, $p$, between 0, 0.5, and 1. 
% We report the following metrics. 

%\begin{itemize}
\noindent\textbf{Alignment.} Alignment measures the proportion of rules aligned with the user's feedback and quantifies the benefit of rule-based summary. The higher it is, the better as it shows that rules reflect the user's views on labeling. 
If a rule $r$ matches 5 warnings and 4 of them have the same label assignment as the user's labeling, we say $r$ is 80\% aligned with the user's labeling. 
We then measure the percentage of \tool{}'s rules that are least $k\%$ aligned. 
We report our results with a target alignment threshold $k$ of 80\%, i.e., an aligned rule is at least 80\% consistent with the user's feedback.

\begin{figure}[!t]
  \centering
  \includegraphics[width=0.9\columnwidth]{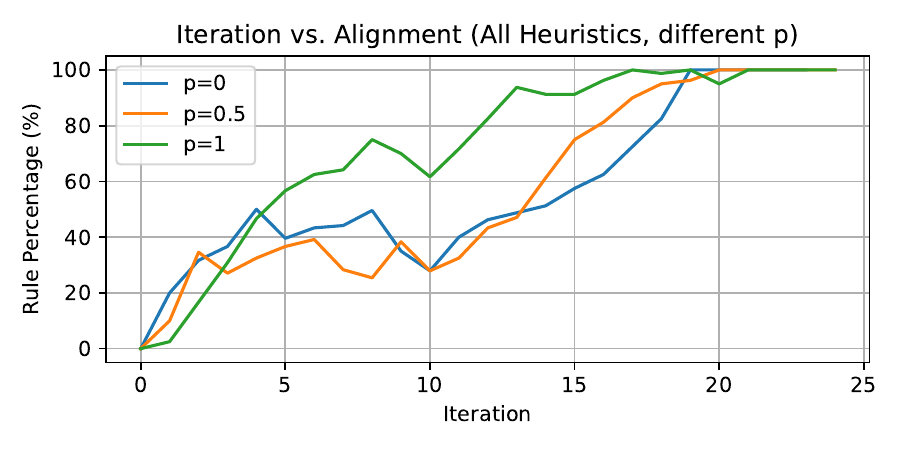}
  \caption{Average alignment results over 20 simulation runs: effect of feedback probability $p$. When providing rule-level feedback ($p=1$), the increase in alignment occurred at earlier iterations compared to providing only feedback to warnings ($p=0$) or less frequent rule-level feedback ($p=0.5$).}
  \label{fig:avg_results_h}
\end{figure}

\begin{figure}[!t]
  \centering
  \includegraphics[width=0.9\columnwidth]{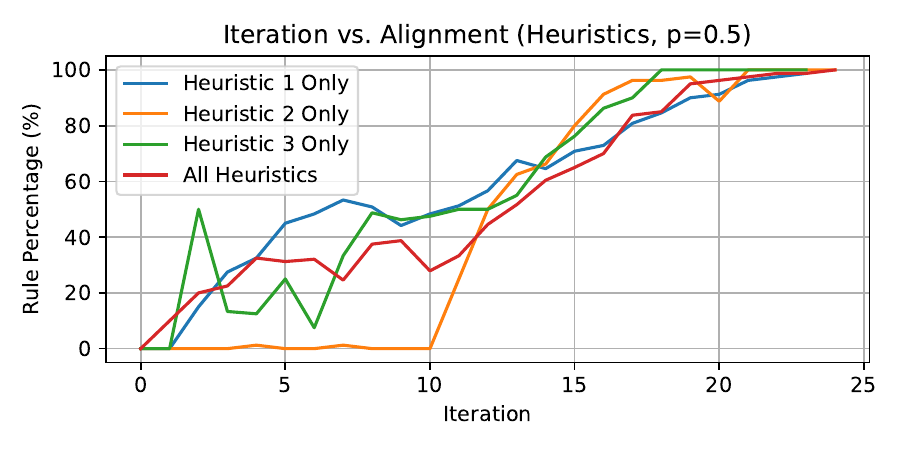}
  \caption{Average alignment results over 20 simulation runs: effect of inspection-order heuristics.}
  \label{fig:avg_results_p}
\end{figure}

 % \caption*{(a) Average alignment (left) and conciseness (right) results for 20 simulation runs over the null pointer dereference warnings found in Alibaba Nacos. \tool{} achieves a high level of alignment and conciseness quickly, independent of the heuristics employed by the simulated user.}

%  \caption*{(b) Average alignment (left) and conciseness (right) results for 20 simulation runs over the null pointer dereference warnings found in Alibaba Nacos. When providing rule-level feedback ($p=1$), the increase in alignment and convergence of conciseness occurred at earlier iterations compared to providing only feedback to warnings ($p=0$) or less frequent rule-level feedback ($p=0.5$).}

\subsection{Results}

Figure \ref{fig:avg_results_h} presents the alignment of the rules as the user iteratively interacts with \tool{}.
Providing feedback at the warning level only ($p=0$) frequently triggers rule refinement and produces noticeable fluctuations in alignment. In contrast, rule-level feedback ($p=1$) yields a much smoother alignment. When providing rule-level feedback only ($p=1$), the alignment score reaches 80\% ($k=80\%$) within 11.8\textsuperscript{th} interaction, the first time. When providing instance-level feedback only ($p=0$), the alignment score reaches 80\% the first time, within 17.8\textsuperscript{th} interaction. 

Next, we investigate the effect of varying user behavior. We ran 20 simulations, while varying user behavior 
% in terms of the examination order 
(Heuristic 1: shorter warnings first, Heuristic 2: similar API calls first, Heuristic 3: similar container names first, and 4: all three heuristics).  
%Figure \ref{fig:avg_results_p} shows the results for different heuristics at $p=0.5$, and Figure \ref{fig:avg_results_h} reflects the results for Heuristic 4, varying the values of $p$. 
Figure \ref{fig:avg_results_p} shows that under different user behavior, rules become aligned with user feedback within 18 iterations. Inspecting warnings in the order of short length or based on the container names leads to more stable and smooth alignment improvement.   
% Here we cannot clearly state the superiority of any heuristic to another. It could be argued that the different heuristics that mimic the user behavior in our user study result in aligned inferred rules to user subjectivity.
%Figure \ref{fig:avg_results_p} shows the impact of providing more frequent rule-level feedback on the the alignment of the inferred rules. Providing more frequent rule-level feedback leads to faster alignment. 
%to 100\% rule alignment. Likewise, we observe a similar trend for conciseness, which had greater stability at $p=1$ than at $p=0.5$ and $p=0$. 

%$p=1$ results in a faster convergence compared to $p=0.5$.
% Similarly, when $p=0.5$, a faster convergence is still seen in contrast to the case where no rule-level feedback is used ($p=1$). 
% We analyze conciseness to observe if there is a correlation between the inferred rules and the uninteresting warnings . 
% A higher correlation should result in fewer fluctuations and more stability in the conciseness results over the iterations. 

% This shows less alignment of warnings labeled uninteresting, and summary rules trying to match uninteresting warnings.
% \hj{What's the takeaway?}

%\hj{Can we say something about how the use of rules doesn't lead to labels different from the individual inspections by the users?}
% Miryung's answer to HJ. I tried to emphasize 'alignment' results

\vspace{-1mm}
\begin{tcolorbox}[colback=black!5!white,colframe=white!50!black]
\tool{}'s rules become aligned with user feedback as interaction progresses. Providing rule-level feedback results in faster alignment than providing warning-level feedback. 
\end{tcolorbox}
\vspace{-2mm}

%for warnings of null pointer dereferences and resource leaks

\section{Discussion}\label{sec:discuss}
\textbf{Inferring configurable and personalized rules.}
A high degree of variation exists among the rules formulated by participants with the help of \tool{}. This aligns with our design goal: \tool{} emphasizes grouping related warnings for sensemaking, not filtering them away. \tool{} achieves this by creating personalized rules based on each user's feedback.
Accordingly, learned rules aim to maximize the concentration (and coverage) of similarly labeled instances within each group (e.g., many ``Uninteresting''), yet do not require complete purity; mixed groups are acceptable and gradually refined as feedback accrues. In the post-study survey, participants indicated a preference for highlighting code expressions over pre-defined predicates (i.e., package and class names, subtyping, used fields, called methods, and return type), showing that participants appreciated the interactive features of \tool{}.
Several participants also asked for even more customization, including defining rules from scratch. Future work can enhance more customization by enabling direct editing of rules with dual modalities of logic rules and natural-language descriptions.
%\burak{Added emphasis on \tool{}'s aim on grouping and NOT filtering.}

In the current state of rules inferred by \tool{}, we observed positive feedback as noted by P8: \textit{``These [Rules] help to avoid human bias when seeing the same repeated structure again. It eases the look up task for similar warnings from the same packages and classes thus narrowing down the search space.}'' and by P9: \textit{``This [Rule inference] is critical to the inductive reasoning in the tool.''} However, some participants noted they preferred more granular rules, as noted by P6: \textit{``The package containment rules didn't seem to help since the package rule was often too broad \texttt{(e.g. org.*)}.''}

%\burak{Added particpant feedback on rules. It is important to note that the rules were found useful 3/5. Their verbal feedback indicated that most of the time they actually did not find the rules inferred by \tool{} were not too useful. Column D: \url{https://docs.google.com/spreadsheets/d/1Xuom5tSTPl8xW_rzpMZikRIONRENcxlQXM6x4uo_HH8/edit?resourcekey=&gid=1230562121#gid=1230562121}}

\noindent\textbf{Sensemaking quality.}
A key finding of our user study is that \tool{} improved participants' ability to articulate commonalities among uninteresting warnings. Participants listed 13\% more common symptoms (1.53 vs.~1.35) and wrote longer, more detailed descriptions (4.6 vs.~3.1 sentences on average) when using \tool{} compared to the baseline. This suggests that interactive summary rules do not merely organize warnings superficially, but enable developers to construct richer mental models of warning patterns. The ability to articulate commonalities is directly actionable: developers can use these insights to de-prioritize groups of similar warnings, update inspection policies, or communicate rationales to teammates.

\noindent\textbf{Developer confidence.}
Participants reported greater confidence when using \tool{} (3.4 vs.~2.6 on a 5-point scale). This is notable given that static analysis tools are often perceived as generating too many false positives, leading to skepticism and tool abandonment~\cite{johnson2013don}. By helping developers recognize patterns among uninteresting warnings, \tool{} may help rebuild trust in static analysis tools by making the inspection process feel more manageable and purposeful.

\noindent\textbf{Efficiency of inspection.}
While \tool{} increased user confidence, some participants took longer to examine tool-generated warnings. 
On average, participants using the baseline inspected over 20\% more warnings.
Having to review the rules carefully slowed down the participants, which may indicate the care taken to reason about the commonalities among related warnings. P6 mentions ``\textit{efficiency gained} [using \tool{}] \textit{was offset by the time I spent trying to get the inference engine to generate the rules I wanted}''. 
% While this was the prominent case for our limited task of 25 warnings, we view that in a real-life scenario, the same task will contain considerably more warnings, which will enable showing the advantage of grouping rules, with larger number of warnings accumulated in each group more compared to the baseline task.

\noindent\textbf{Utility of individual features.}
Among \tool{}'s features, the warning distribution display was rated most useful (3.8/5, 11 out of 14 participants), followed by highlighting code expressions (3.7/5, 12 out of 14). These ratings suggest that both seeing how many warnings a rule covers and providing fine-grained code-level hints are central to the sensemaking workflow. The label-all feature was more situational, found useful by 8 out of 14 participants — most effective when participants were confident that a rule captured a coherent group of warnings.

\noindent\textbf{Implications of the simulation.}
The simulation study shows that rule-level feedback leads to faster alignment than instance-level feedback (11.8 vs.~17.8 interactions to reach 80\% alignment). This has a practical implication: in real workflows where warning sets can be large, the ability to confirm or reject an entire group at the rule level substantially reduces the number of interactions needed to align \tool{}'s groupings with the developer's intent. This suggests that rule-level feedback is not merely a convenience feature, but a key driver of \tool{}'s efficiency at scale.

\noindent\textbf{Ease of use.}
There is an initial learning curve for using \tool{} compared to the examination of individual warnings.
% This is unsurprising since using \tool{} requires an initial learning curve.
% switches up the warning inspection strategies of the participants, who  reported examining warnings one-by-one in their previous experience with static analysis tools.
% Moreover, we argue that \tool{}, in comparison to the baseline, which has no features has a certain learning curve caused by it having various components, that may have not been grasped thoroughly by the participants given the limited amount of time.
Nevertheless, tools perceived to be difficult to use can still provide 
% utility and
value~\cite{sarkar2023should}.

\noindent\textbf{Failure Cases.} \tool{}'s rule inference algorithm is designed to group warnings based on the predicates present in self-contained functions and consistent user labeling. In the case of highly heterogeneous warning patterns with commonalities that span multiple procedures (e.g., taint flows or data races), or inconsistent user labeling, \tool{} may produce lower-quality rules.

\noindent\textbf{Threats to validity.}
One threat to validity is related to the choice of bug types (i.e., analysis kind) used in our study. We focused on null pointer dereferences and resource leaks since these kinds of analysis are widely supported by commercial static analyzers.
Grouping based on containment, type hierarchy, code similarity, and API methods and fields used in \tool{} should generalize to grouping other kinds of tool-generated warnings on code snippets. 
A second threat concerns the 10-minute cap per task. Short windows may bias participants toward shallow triage. We mitigated this by providing a tutorial and warm-up, using identical time limits for both conditions, and counterbalancing tool/order to control learning effects; still, longer or longitudinal use may reveal additional benefits. %\burak{Added 10-minute task time as a threat to the validity of our user study evaluating \tool{}.}
While our study included only 14 participants, studies that use a within-subject study with a crossover design require fewer participants~\cite{chasins2021pl}, as the crossover minimizes variability~\cite{vegas2015crossover}.
% \sout{Moreover, our participants had a wide range of programming experiences, and included both graduate students and professional developers.}

\section{Related Work}\label{sec:related}
\textbf{Sensemaking.} Prior sense-making systems (Selenite and Synergi for large text corpora~\cite{selenite, synergi}, Mesh for e-commerce trade-offs~\cite{mesh}, Unakite for programming Q\&A snippets~\cite{unakite}, and decision-support tools such as Dreamsheets, Pail, and cost-structure models~\cite{dreamsheets, beyond_code_gen, cost_structure}) show that exposing similarities and contrasts accelerates insight, 
% yet they sit largely outside day-to-day software-engineering practice. 
but are not applied to software engineering tasks.
The closest work~\cite{confidence_highlighting,uncertainty_high,calibration_of_correctness} highlights confidence and uncertainty cues 
% but are limited to inspecting 
only on 
a single code artifact at a time. 
\tool{} enables variation-driven sensemaking for triaging static-analysis warnings. 

\textbf{Filtering and reprioritizing.}
A large amount of work has focused on post-processing  warnings generated from static bug finding tools~\cite{kim2007prioritizing,hanam2014finding,heckman2009model,ruthruff2008predicting,shen2011efindbugs,wang2018there,kang2022detecting,yedida2023find}, which filter or re-prioritize generated warnings so that the uninteresting findings should appear last. 
Some of this work ~\cite{wang2018there,yedida2023find,kharkar2022learning,yang2021understanding} have shown that machine learning techniques could be effective at predicting which warnings are likely to be uninteresting. 
However, 
% machine learning techniques have two key limitations: they rely on pre-labeled training data, necessitating initial human inspection, and 
they fail to support developers for sensemaking the warnings.
This can be helpful, but 
does not 
% learn from developers' individual preferences or 
help developers establish accurate mental models to understand the alerts. 
% In contrast, SWIRL treats inspection as an interactive sensemaking activity. SWIRL induces logic rules that emphasizes commonalities e.g., containment, subtype relations, shared API calls and can be iteratively refined.
% based on user's feedback. 
The goal is not to bury uninteresting warnings, but to synthesize rules for developing insights and communicating rationales.
% to teammates, 
% and codify organization-specific suppression policies.

\textbf{Clustering.}
Prior work on clustering static-analysis warnings identifies cause points (program locations whose imprecision triggers many alerts (e.g., a method spuriously deemed reachable) and employs interactive workflows to reduce triage effort~\cite{muske2022,dillig2012automated,muske2016cause,zhang2017effective}. 
However, warnings are ignored for reasons other than analysis imprecision.
For supporting sensemaking, \tool{} surfaces shared characteristics beyond analysis-imprecision program points: users can group related warnings by containment (e.g., package/class), by salient code expressions (e.g., specific API calls), and by their composition. 
% \tool{} also supports iterative feedback at both the rule level and the instance level.
%\burak{Added another paragraph about how sensemaking differs from FP/FN classification}

\textbf{Manual inspection of tool-generated warnings.}
Other studies have proposed methods of inspecting warnings.
Some analyzers focus on presenting proposed fixes~\cite{barik2016quick} together with generated warnings. Mangal et al.~present interactive static analysis, allowing users to make tradeoffs in the approximation made by the static analyzer~\cite{mangal2015user}.  
Buckers et al.~allow users to compare the warnings generated by different analyzers~\cite{buckers2017uav}.
% In contrast, \tool{} specifically 
However, they do not 
support the sensemaking process necessary  for developers to provide accurate feedback.
% , and for doing so.
% of winnowing out uninteresting warnings, using their common characteristics.

Ma et al.~\cite{ma2019reorganizing} categorize warnings by their bug types and the source and sink function calls. 
Unlike \tool{}, these categories (similar sources and sinks) are determined prior to users' inspections, and are not updated on-the-fly following the user's gradual and interactive feedback. 
Compared to \tool{}, the warnings are categorized without common code expressions at a fine granularity.
Muske et al.~\cite{muske2016cause} identify the cause points behind imprecise analyses.
Their technique heuristically detects locations that are challenging for static analysis before posing queries to the user based on these locations.
In contrast, \tool{} does not depend on pre-defined assumptions about what locations are challenging for static analysis, and can learn custom rules about the implicated code snippets with interactive user feedback. 
% We found in our user study that the perceived utility of warning aspects, including containment relationships and program constructs, was subjective and differed across users.

\textbf{Code pattern inference using human feedback.}
Studies have developed methods of inferring code patterns using human feedback~\cite{kang2021active,zhang2015interactive,sivaraman2019active,wang2023synthesizing,kang2024scaling,garg2022synthesizing},
 % for code search~\cite{sivaraman2019active,wang2023synthesizing,kang2024scaling},
% and synthesizing rules for code quality~\cite{garg2022synthesizing}.
% Several techniques employ human feedback to improve or refine a code  pattern~\cite{kang2024scaling,zhang2015interactive}.
Similar to these approaches, \tool{} includes a rule-learning component, but composes rules over two different modalities: (1) code organization in terms of containment, subtyping, subclassing, called methods, used fields, return type relationships and (2) code expressions harvested with interactive highlighting feedback for fine-granular rules.

% {\tool} is orthogonal to approaches that find similarities in Abstract Syntax Trees~\cite{sousa2021learning,serrano2020spinfer}. Those tools match warnings based on a single code modality, whereas {\tool} matches warnings using a combination of two modalities: code and containment relationships. This was supported by our user study, where participants were better able to articulate commonalities between warnings using rules from both modalities.

\textbf{Sensemaking of data.}
Many studies have proposed tools for sensemaking of large collections of data.
% \sout{Several approaches learn rules to analyzing data.}
Tempura~\cite{wu2020tempura} learns structural template for grouping textual data. 
\tool{} has similarities to PaTaT~\cite{gebreegziabher2023patat}, a tool that learns patterns based on user-annotated codes as the user inspects data.
These tools support human subjectivity and agency~\cite{jiang2021supporting} in assigning labels to data, but both Tempura and PaTaT focus on textual data, while \tool{} is designed for inspecting warnings on code snippets reported by bug finding tools.

OverCode~\cite{glassman2015overcode}, Examplore~\cite{glassman2018visualizing}, and ExampleNet~\cite{yan2021visualizing} summarize a collection of code by providing a bird-eye view or a structure over variations in code.
% \sout{They summarize code that are similar in how they address the same programming task, use the same API, or use deep learning models respectively.}
Instead of summarizing the variations in code, 
{\tool} highlights common patterns of tool-generated warnings by leveraging containment relationships and common code expressions, and enable refinements through interactive feedback.

% \textbf{Interfaces for inspecting and debugging code.}
% \tool{} is an interface for debugging static analysis warnings.
% WhyLine~\cite{ko2004designing} is an interface for programmers to ask questions regarding a program's behavior and to view answers in terms of runtime executions.  
% Its goal is orthogonal to \tool{}, as it focuses on debugging a single program, as opposed to constructing abstraction rules for static analysis results.  

% \tool{} focuses on 

% \tool{} visualizes a code pattern inferred from the warnings. 
% This resembles Examplore~\cite{}, which summarizes code about API usages on GitHub by overlaying common code elements.
% \tool{}'s code pattern aims to represent the root cause of false alarms instead of visualizing common API usages.

% \miryung{finished up to here. hongjin, i remember there were some related work we should cite based on our earlier ICSE 2025 feedback. can you double check reviews?}
% \hj{what we missed was a more detailed discussion of Ma et al.~\cite{ma2019reorganizing}, which i expanded on above, and Revisar}

\section{Conclusion}\label{sec:conclusion}
In this paper, we propose \tool{}, which provides systematic support for personalized sensemaking of tool-generated warnings by harnessing user feedback with active learning.
Our key insight is that automatically refined summary rules can enable users to contrast common characteristics among related warnings and that users' specific hints should be incorporated to re-infer summary rules.
% These rules group related warnings using predicates capturing containment, subtyping, method call, field usage relationships and the use of similar code expressions highlighted by users, allowing users to formulate custom abstraction about which warnings should be (or should not be) acted upon.

We evaluate \tool{} with two different evaluation methods. First, our user study reveals that participants developed individualized, custom rules for grouping related warnings, confirming our insight that one-size-fits-all policy may not work during the cognitive, sensemaking process and that users desire configurability. 
% \sout{\tool{} helps users discern commonalities and differences among different groups of uninteresting warnings.} 
When using \tool{}, users reported lower levels of frustration, higher levels of confidence, and could articulate more details about underlying similarities regarding uninteresting warnings.  Second, our simulation study found that, by enabling users to provide feedback at the level of rules, \tool{}'s resulting rules become aligned faster with labeling feedback. 

\section{Data Availability}\label{sec:data}
The replication package and study's data are available at \url{https://github.com/UCLA-SEAL/SWIRL}.

\section*{Acknowledgments}
This work is supported by the National Science Foundation under grant numbers 2426162, 2106838, and 2106404. It is also supported in part by funding from Amazon and Samsung.

% ====================
% Bibliography
% ====================
\balance
\bibliographystyle{IEEEtran}
\bibliography{IEEEabrv,interactive_static_analysis}

\end{document}